# On the Dynamical Complexity of Small-World Networks of Spiking Neurons

Murray Shanahan

Department of Computing, Imperial College London, 180 Queen's Gate, London SW7 2AZ, UK.

**Abstract**

A computer model is described which is used to assess the dynamical complexity of a class of networks of spiking neurons with small-world properties. Networks are constructed by forming an initially segregated set of highly intra-connected clusters and then applying a probabilistic rewiring method reminiscent of the Watts-Strogatz procedure to make inter-cluster connections. Causal density, which counts the number of independent significant interactions among a system's components, is used to assess dynamical complexity. This measure was chosen because it employs lagged observations, and is therefore more sensitive to temporally smeared evidence of segregation and integration than its alternatives. The results broadly support the hypothesis that small-world topology promotes dynamical complexity, but reveal a narrow parameter range within which this occurs for the network topology under investigation, and suggest an inverse correlation with phase synchrony inside this range.

I. INTRODUCTION

The existence of both functional and structural networks with small-world properties in the brains of a variety of animals is now well established [1, 2], and the evolutionary, metabolic, and computational constraints likely to favour neural networks with small-world topologies have been the subject of much recent discussion [3, 4]. At the same time, the theme of dynamical complexity – the extent to which both segregated and integrated activity are present in the same system – has become important in the attempt to understand how sophisticated cognition emerges from the activity of large numbers of neurons [5, 6, 7]. In pursuit of this theme, Sporns, *et al*. [8] report a computer experiment with an evolutionary algorithm in which an initial population of randomly connected networks evolved towards small-world topologies when subject to a selection criterion that favoured dynamical complexity (according to one of several formal measures of that concept). Further support for the thesis that small-world topology promotes dynamical complexity in neural networks is provided by Roxin and colleagues, who describe the simulation of a specific class of small-world networks of integrate-and-fire neurons with constant conduction delays [9, 10].

The class of networks investigated by Roxin and colleagues are constructed using the well-known Watts-Strogatz procedure [11], in which the short-range connections in a ring lattice are replaced by long-range connections with a given probability $p$. In [9], these authors demonstrate a variety of different behaviours for their spiking neuron networks depending on $p$. In particular, they show that if $p$ lies within a certain critical range, the resulting networks tend to generate self-sustaining activity. They also show that, given a sufficiently long conduction delay, higher values of *p* within this range support a form of chaotic behaviour which is suggestive of increased dynamical complexity. As Riecke, *et al*. point out [9], the network topology generated by the Watts-Strogatz procedure is a reasonable approximation of a mid-sized patch of cortex that comprises abundant short (less than 100μmm) connections among nearby neurons with proportionally fewer longer-range projections to sites several millimetres away. However, there is strong evidence that cortical anatomy also enjoys small-world properties on a more global scale [1, 2], where it may be characterised as a network of anatomically distinct regions that are densely intra-connected at the level of grey matter and interact with each other through relatively sparse long-range white matter projections [12]. This kind of small-world topology is not well modelled by the Watts-Strogatz procedure, which produces networks whose density of short-range connections is uniform, and in which markedly segregated areas therefore do not exist.

The present paper reports an experiment whose aim was to deepen our understanding of the relationship between dynamical complexity and connectivity in so-called "modular" small-world networks [13, 14]. These networks, which comprise a number of densely intra-connected clusters linked by a small set of inter-cluster connections, are better approximations of cortical structure at a global scale than those produced by the Watts-Strogatz procedure. However, echoing the Watts-Strogatz procedure, the network construction method employed here includes a probabilistic rewiring phase to establish inter-cluster connections, and it can therefore be parameterised by the rewiring probability *p*. So in the experiment *p* was used to capture each network's structure, along with its small-world index. Causal density – which assesses the extent to which each variable in a system exercises independent influence on the others – was the chosen measure of dynamical complexity since, as well as being able to distinguish between individual and systemic dependencies in multivariate data, it is sensitive to temporally smeared evidence of segregation and integration thanks to its use of lagged observations [5]. This feature turns out to be essential for detecting the complexity trend that is the main result of the paper.

In line with the findings of Sporns *et al*. [8] and Riecke, *et al*. [9], we would expect dynamical complexity to correlate positively with small-world index. Intuitively, a high small-world index (assuming sparse connectivity) indicates that distinct portions of the network enjoy considerable structural segregation yet remain significantly interconnected, and it is reasonable to expect this balance to manifest in the dynamics as loosely coupled islands of largely independent activity. The results of the experiment broadly support this hypothesis. However, contrary to expectations, they show that a high small-world index does not guarantee dynamical complexity. Rather, there is a narrow parameter range within which the balance in question is struck, because networks with too much inter-cluster connectivity, though still enjoying a high small-world index, tend towards entrainment, a dynamical regime which is highly integrated but lacks segregation. High dynamical complexity only occurs with sparsely inter-connected clusters, although this is indetectable without the use of lagged observations because such networks only exhibit integration over time. Since dynamical complexity is a plausible prerequisite for flexible cognition this suggests that the structural connectivity matrices of large-scale cortical networks will conform to a tighter set of constraints than was previously realised.

II. THE NETWORK AND ITS CONSTRUCTION

Each randomly generated network in the experiment comprised 1000 spiking neurons with variable conduction delays, of which 800 (80%) were excitatory and 200 (20%) inhibitory. The spiking neuron model used was that of Izhikevich [15]. The model is defined by the following three equations.

$$\dot{v} = 0.04v^2 + 5v + 140 - u + I \quad (1)$$

$$\dot{u} = a(bv - u) \quad (2)$$

$$\text{if } v \geq 30 \text{ then } \begin{cases} v \leftarrow c \\ u \leftarrow u + d \end{cases} \quad (3)$$

where $v$ is the neuron's membrane potential, $I$ is its input current, and $u$ is a variable that regulates the recovery time of the neuron after spiking. Eqn. (3) describes the way the neuron is reset after spiking, which is assumed to occur when its membrane potential reaches 30mV. Following Izhikevich [15], the parameters of the neuron model were assigned non-uniformly to endow the population with a variety of signalling behaviours. For excitatory neurons, the values used were $a = 0.02$, $b = 0.2$, $c = -65 + 16r2$, and $d = 8-6r2$, where $r$ is a uniformly distributed random variable in the interval [0,1]. For inhibitory neurons, the values used were $a = 0.02 + 0.08r$, $b = 0.250-0.05r$, $c = -65$, and $d = 2$, with $r$ as above.

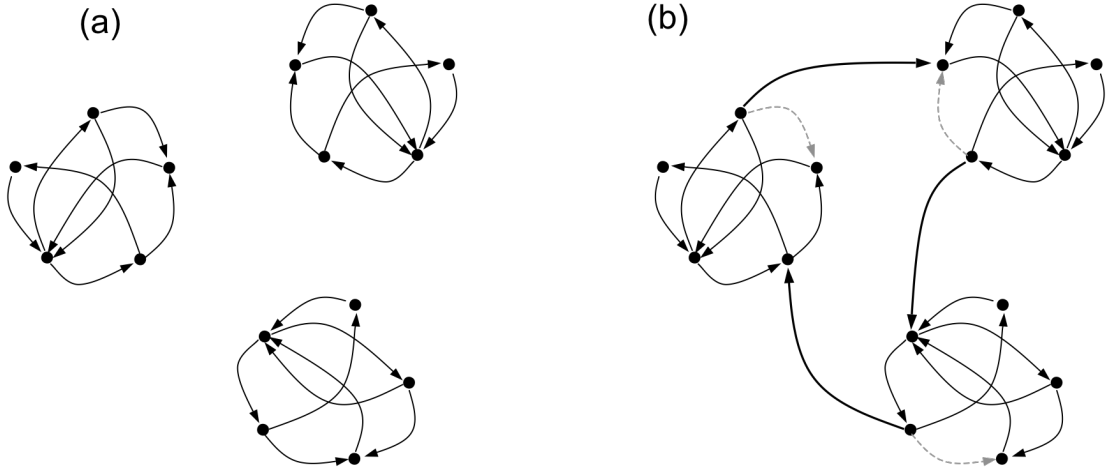

FIG. 1: Small-world network construction. The first phase of wiring (a) produces a set of separate but densely intra-connected clusters. In the second phase (b), a small subset of the connections established in the first phase is replaced with "long-range" inter-cluster connections.

Consider a time $t$ and a neuron $i$, and let $\Phi$ be the set of all neurons $j$ that fired at time $t-\delta$ where $\delta$ is the conduction delay from neuron $j$ to $i$. Then the input current $I$ for neuron $i$ at time $t$ is given by:

$$I(t) = \sum_{j \in \Phi} S_{i,j} F \qquad (4)$$

where $S_{i,j}$ is the synaptic weight of the connection from neuron $j$ to $i$ and $F$ is a scaling factor. $F$ was set to 30 in all the experiments described here.

The wiring regime was as follows (Fig. 1). Every excitatory neuron had 20 synaptic connections, of which 16 (80%) were to other excitatory neurons and 4 (20%) were to inhibitory neurons. Synaptic weights were randomly assigned, drawn from a uniform distribution over the interval [0,0.7] for excitatory connections and from a uniform distribution over the interval [-2,0] for inhibitory connections. Two experiments were carried out. In the first, the 800 excitatory neurons were organised into 8 clusters of 100, while in the second they were organised into 10 clusters of 80. Excitatory connections were established using a two-phase procedure reminiscent of the Watts-Strogatz method. In the first phase, for every excitatory neuron in the network 16 connections were made to randomly chosen excitatory neurons within the same cluster. Conduction delays in the range 1ms to 20ms were assigned randomly to each connection. In the second phase, a rewiring decision was made for every excitatory connection established in the first phase. With probability $p$, an established intra-cluster connection was replaced with a connection to a randomly chosen neuron within a randomly chosen different cluster (preserving the original conduction delay). A method for constructing a very similar class of networks is described by Pan and Sinha [9], but their method does not use a separate rewiring phase.

As with the Watts-Strogatz procedure, the present wiring regime ensures that various statistics are preserved across a population of randomly generated networks whatever the value of $p$, including the number of nodes, the number of connections, the mean synaptic weight averaged over the whole network, and the mean conduction delay averaged over the whole network. Note that if $p = 1$ a fully randomised network results, while if $p = 0$ the resulting network comprises eight disconnected sub-networks. The inhibitory neurons were also organised into eight clusters, one per excitatory cluster, and connections between excitatory and inhibitory neurons within corresponding clusters were established randomly in both directions. Connections to and from inhibitory neurons were unaffected by the rewiring phase.

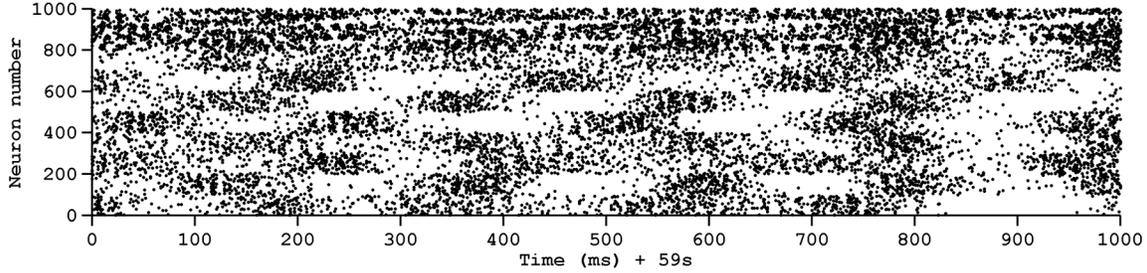

FIG 2. Raster plot of the last 1000ms of a representative trial (number 21). Neurons numbered 1 to 800 are excitatory, and organised into 8 sequentially numbered clusters (1 to 100, 101 to 200, etc.). Neurons numbered above 800 are inhibitory.

III. METHODS AND MEASURES

A series of 1000 trials were conducted: 500 with 8-cluster networks and 500 with 10-cluster networks. For each trial, the probability $p$ was drawn from a uniform distribution over the interval [0, 0.15] and a new network was generated using this value. The resulting network was then run for 60s of simulation time. In each run, a single neuron was forced to spike at $t = 500$ms. Since the neurons received no input current prior to 500ms, the network remained quiescent until then. But following the injection of this spike, a period of self-sustaining network activity ensued. In some runs this activity died out before $t = 60$s, but in many cases (depending on the value of $p$, as we shall see) it lasted for the entire duration of the run. Figure 2 shows a raster plot of all neuron firings during the last second of a representative 8-cluster trial.

The small-world index of each network generated was computed. This is defined as follows. Consider a graph $G$ graph with $n$ nodes and $k$ edges per node on average. The *path length* between any pair of nodes in $G$ is the number of edges in the shortest path between those nodes, and *G*'s *mean path length* $\lambda_G$ is the path length averaged over every pair of nodes in *G*. The *clustering coefficient* of a node *P* in *G* is the fraction of the set of all possible edges between immediate neighbors of *P* that are actual edges, and the *clustering coefficient* $\gamma_G$ of the whole graph *G* is the clustering coefficient averaged over the set of all its nodes. A sparsely connected graph (where $k \ll n$) is said to be small-world if its mean path length is comparable to that of a randomly connected graph with the same *n* and *k* but its clustering coefficient is higher. This property can be quantified as the *small-world index* $\sigma_G$ of *G* [16], which is defined as

$$\sigma_G = \frac{\gamma_G / \gamma_{rand}}{\lambda_G / \lambda_{rand}}$$

where $\gamma_{rand}$ and $\lambda_{rand}$ can be approximated as $k/n$ and $\ln(n)/\ln(k)$ respectively [11].

In addition to the small-world index, an estimate of the causal density of the interactions among the network's eight clusters was computed for each trial [17, 5]. The concept of causal density is based on that of *Granger-causality* [18], which is a measure of the causal influence among the variables of a dynamical system. To grasp the idea of Granger-causality, consider a trio of time series $X_1(t)$, $X_2(t)$, and $X_3(t)$, and suppose $X_1(t)$ is described by the following autoregressive model:

$$X_1(t) = \sum_{j=1}^{m} A_j X_1(t-j) + B_j X_2(t-j) + C_j X_3(t-j) + \varepsilon_{ABC}(t)$$

where *m* (the model order) is the maximum observation lag, *A*, *B*, and *C* are vectors containing the coefficients of the model (indexed by observation lag), and $\varepsilon_{ABC}$ is the prediction error. This can be compared with the following model of $X_1(t)$, in which the $X_2$ term is absent:

$$X_1(t) = \sum_{j=1}^{m} A_j X_1(t-j) + C_j X_3(t-j) + \varepsilon_{AC}(t)$$

Now, $X_2$ is said to *Granger-cause* $X_1$ if the variance of $\varepsilon_{ABC}$ is significantly less than the variance of $\varepsilon_{AC}$, that is to say if the inclusion of the $X_2$ term helps to predict $X_1$. Assuming $X_1$ to $X_3$ are covariance stationary, this significance can be determined using an F-test. Note that for $X_2$ to Granger-cause $X_1$, it must exercise an influence over and above that of $X_3$. Clearly $X_2$ and $X_3$ can be treated symmetrically, and the extension to any number of variables is straightforward. The *causal density* of a system of variables $X_1...X_n$ is then defined as $\alpha/n(n-1)$, where $\alpha$ is the number of pairs of variables $\langle X_i, X_j \rangle$ such that $X_i$ Granger-causes $X_j$. In other words, causal density measures the proportion of all possible causal relations among system variables that is statistically significant.

To see that causal density assesses the antagonistic balance between integration and segregation, and is therefore a valid measure of dynamical complexity as claimed in [5], let's consider how the measure behaves under the two boundary conditions of low segregation with high integration and low integration with high segregation. Consider a system of variables $S = \{X_1...X_n\}$, and suppose the system is poorly segregated but highly integrated. In this case, there will be *many large* subsets of $S$ whose members are highly correlated, and there will be correspondingly few instances of a variable $X_i$ Granger-causing another variable $X_j$. This is because many large subsets of $S$ will be good predictors of a typical $X_j$, entailing that $S$ will generally be no better at predicting $X_j$ than $S-\{X_i\}$. That is to say, an equally good autoregressive model would be possible *with or without* the inclusion of $X_i$. So the causal density will be low.

At the other extreme, suppose the system is poorly integrated but highly segregated. In this case there will be (at most) a *few small* subsets of $S$ whose members are highly correlated. Once again there will be few instances of $X_i$ Granger-causing $X_j$. The reason now is that few (if any) subsets of $S$ will be good predictors of a typical $X_j$, and $S$ will only be better at predicting $X_j$ than $S-\{X_i\}$ if $X_i$ happens to be in one of those rare subsets. So the causal density will again be low. Only when there are numerous cases of $X_i$ influencing $X_j$ independently of the influence of the rest of the system will the causal density be high, and this only occurs if the system is both well segregated and well integrated.

For comparison, an alternative, information-theoretic measure of dynamical complexity was also computed for each trial, which approximates its "neural complexity" [7]. According to Tononi, *et al.* [7], the "neural complexity" C of a system of variables $S=\{X_1...X_n\}$ can be estimated by

$$C(S) = \sum_{i=1}^{n} \mathrm{MI}(X_i; S-\{X_i\}) - \mathrm{I}(S)$$

where MI(*X*,*Y*) is the mutual information of *X* and *Y*, and I(*S*) is the *integration* of *S* defined by

$$I(S) = \sum_{i=1}^{n} H(X_i) - H(S)$$

where $H(X)$ is the entropy of $X$.

IV. RESULTS

In each run, firing data was gathered for the network's 800 excitatory neurons, resulting in an 800 by 59000 binary matrix, each element of which represents the occurrence or non-occurrence of a spiking event for the relevant neuron at the specified time point (the first 1000ms of each run was ignored). It would be infeasible to apply a causal density analysis directly to so much data. But knowing the large-scale topology of the network we can reduce it to a small set of time series, one per network cluster. In practice, each such time series was obtained by calculating the mean number of firings per neuron in the cluster per millisecond over a moving 50ms window sampled at 20ms intervals, yielding 8 or 10 time series (depending on the number of clusters) each of length 2950. For causal density analysis to be valid, the time series to which it is applied must be covariance stationary. This can be established using the Augmented Dickey-Fuller (ADF) test. A raw time series obtained by the above method invariably fails the ADF test. So first-order temporal differencing was deployed, generating a new set of 8 (resp. 10) time series of length 2950, in effect representing the change in mean firing rate over time. Since 100% of the differenced time series passed the ADF test in the 8-cluster experiment and 99.75% passed the test in the 10-cluster experiment, these were the subject of the causal density analysis, which was carried out using a model order of 10.

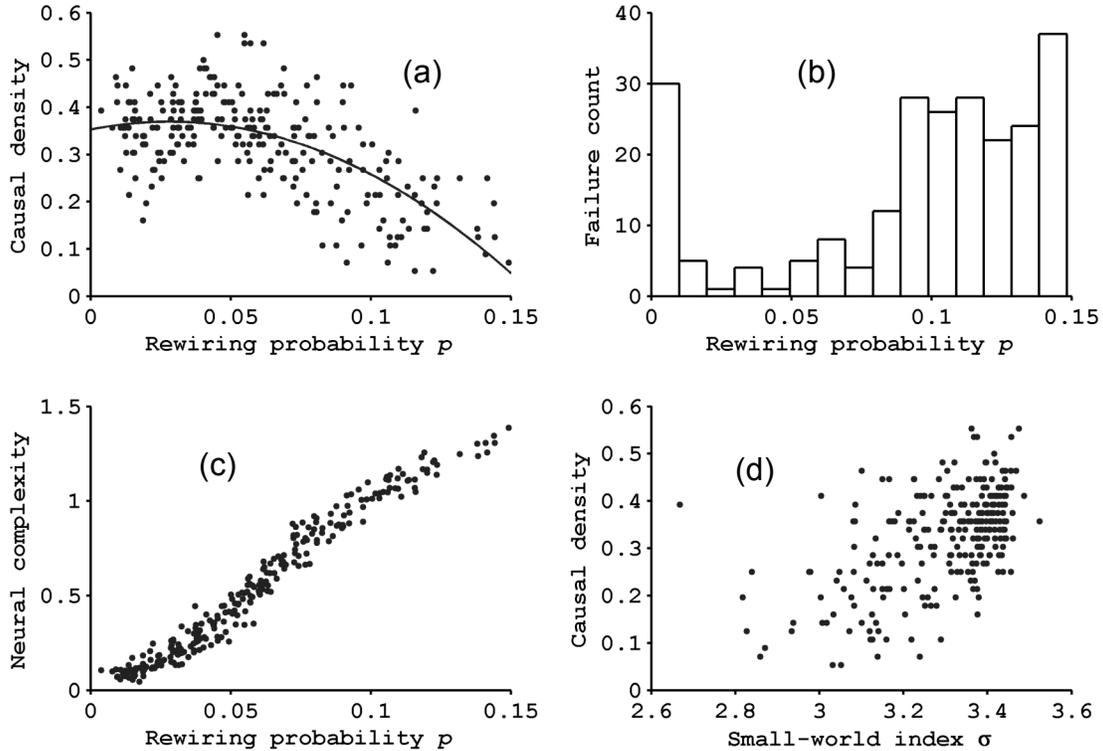

FIG. 3: The relationship between rewiring probability, causal density, neural complexity, failure count, and small-world index for the 8-cluster experiment. See text for details.

The results for the 8-cluster experiment are summarised in Fig. 3. Causal density peaks at around 0.38, when the rewiring probability *p* is approximately 0.05 (Fig. 3(a)), and tails off for values of *p* greater than 0.05. Note that causal density is only assessed for runs that exhibit sustained activity for the full 59.5s following the introduction of the initial spike at 500ms. A run in which this is not achieved is deemed a "failure". There were 265 successful runs out of 500 trials in the 8-cluster experiment. Causal density was not computed for failed runs. The absence of data points close to the Y-axis in Fig. 3(a) indicates that networks generated with *p* less than 0.01 are very rarely capable of sustained activity for 60s. The leftmost column of the histogram in Fig. 3(b) confirms that although 30 networks were generated with p ≤ 0.01 they all produced failed runs.

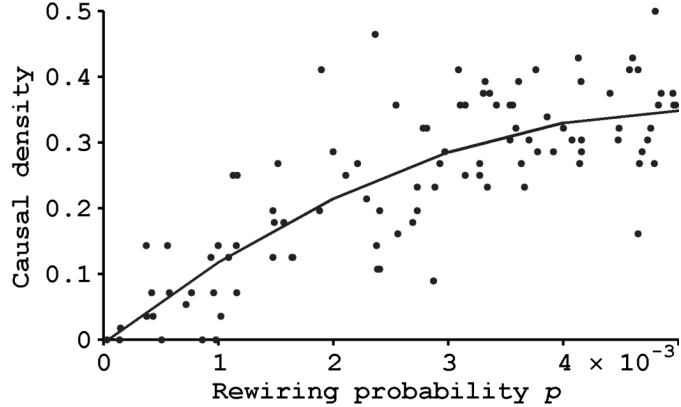

FIG. 4: The relationship between rewiring probability and causal density for low values of $p$. To eliminate failed runs, a single spike was injected to restart activity whenever it died out.

Moreover, as Fig. 3(b) shows, the proportion of networks capable of sustained activation decreases with p > 0.05. In other words, when $p$ falls outside a certain narrow range, from approximately 0.01 to approximately 0.09, it typically generates networks that either fail or have low causal density.

Fig. 3(c) shows the contrasting results of computing an approximation to the "neural complexity" of the same data. With hindsight (see discussion below) it is no surprise that this measure fails to see significant complexity in runs with high causal density, because it is insensitive to integration or segregation that is smeared over time. Only when activity in the clusters begins to synchronise does it detect complexity, which is exactly when the *dynamical* complexity of the system is starting to tail off according to causal density. (A different trend for neural complexity is reported by Buckley & Bullock [19], using a static complexity analysis instead of time series data. Although one of the network topologies they explore is superficially similar to the one used here, their probabilistic rewiring method is in fact different.)

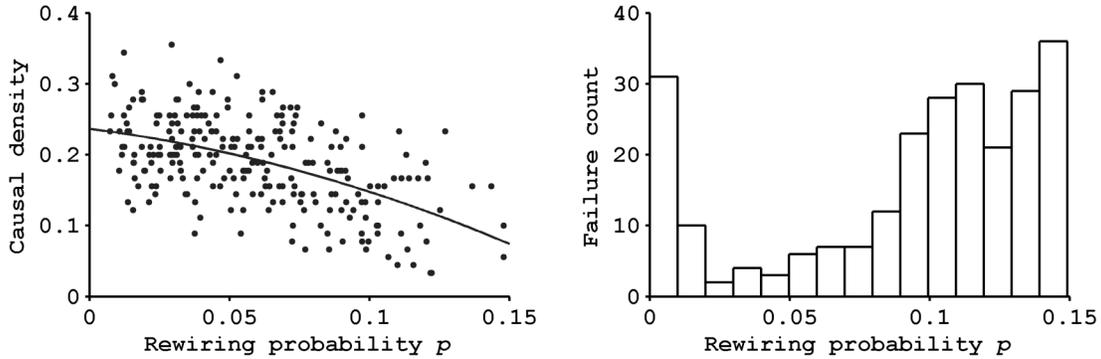

FIG. 5: The relationship between rewiring probability, causal density, and failure count for the 10-cluster experiment. As with the 8-cluster experiment, causal density tails off as rewiring probability increases.

Fig. 3(d) suggests that networks with higher small-world indices ($> 3.3$) give rise to runs with greater causal density ($> 0.25$). (Although this trend is obscured by considerable trial-to-trial variation, it should be borne in mind that the number of data points is much larger for higher small-world indices because these yield fewer failed runs.) However, the small-world index for networks generated with $p > 0.1$ is still high (significantly greater than 1). Fig. 4 presents the results of a variation on the 8-cluster experiment which was designed to investigate very low values of $p$. In this version, to overcome the problem that very low values of $p$ almost always generate failed runs, a single spike was injected to restart activity whenever it died out. The results of the experiment (for 100 trials) reveal a sharp rise in causal density between $p = 0$ and $p = 0.005$ which is invisible in Fig. 3(a) due to the scale and the prevalence of failed runs. This is consistent with the expectation that a network whose clusters are almost disconnected, and which therefore generates highly segregated activity with negligible integration, should exhibit low dynamical complexity. Finally, Fig. 5 summarises the results for the 10-cluster experiment, which yielded 251 successful runs. As with the 8-cluster experiment, these show that causal density tails off as $p$ increases, and that self-sustained activation is most likely to occur when $p$ is within a narrow range (close to 0.04).

To the extent that causal density is acceptable as a measure of dynamical complexity these findings support the conclusion that, for the class of networks under consideration, a sufficiently high small-world index promotes both self-sustained activation and complexity. However, they also indicate that networks that are small-world but that enjoy only modest small-world indices tend not to be capable of self-sustained complex behaviour. There are several avenues of further investigation. For example, the present results were obtained with a relatively large current scaling factor $F$, which causes each cluster in a network to be, so to speak, on a hair trigger. A single spike is typically sufficient to initiate a chain reaction of firing. Consequently very few connections between clusters are required to support complex interactions. To obtain results for lower values of $F$ would require more neurons and synapses, and perhaps other modifications to the experimental setup. Further work is also needed to understand more fully the relationship between dynamical complexity and cluster count, and to gain a proper grasp of the relative merits of causal density and other proposed measures, such as "neural complexity".

V. DISCUSSION

Some insight into the underlying reasons for the reported results can be gained by looking into the rhythmic behaviour of the networks. Visual inspection of the raster plots shows that the level of activation within each cluster is highly rhythmic, waxing and waning at a frequency of approximately 4Hz (corresponding to the theta band in EEG terms). This is to be expected because, due to their dense connectivity, there is a tendency for activation to spread rapidly within a cluster, causing all its neurons to hyperpolarise together. So, within a given cluster, periods of intense activation tend to be followed by periods of total quiescence while all the neurons recover in concert. But thanks to their connections to other clusters – the very connections that confer small-world topology on the network – there is always a significant number of incoming spikes during these quiescent periods which can reignite activity in the cluster when it has recovered. A similar phenomenon is reported by Roxin, *et al*. [10].

Although theta-band activity is common to networks exhibiting both high and low causal density, a visual comparison of typical raster plots suggests a difference in their phase characteristics. In particular, one hallmark of networks with high causal density is the absence of a stable phase difference between the theta waves generated by each cluster, while clusters in networks with low causal density are far more likely to entrain, producing synchronous theta waves. A plausible explanation for this is that an excess of inter-cluster connections (produced with high values of *p*) leads to such strong coupling between clusters that the phenomenon of simultaneous hyperpolarisation extends across cluster boundaries. This causes the clusters to become entrained [20], making each cluster a good predictor for activity in every other cluster and bringing down the causal density accordingly. If this coupling is too strong, the entire network can become quiescent simultaneously, causing self-sustained activation to be extinguished entirely, a frequent occurrence with high values of $p$ (Fig. 3(b)).

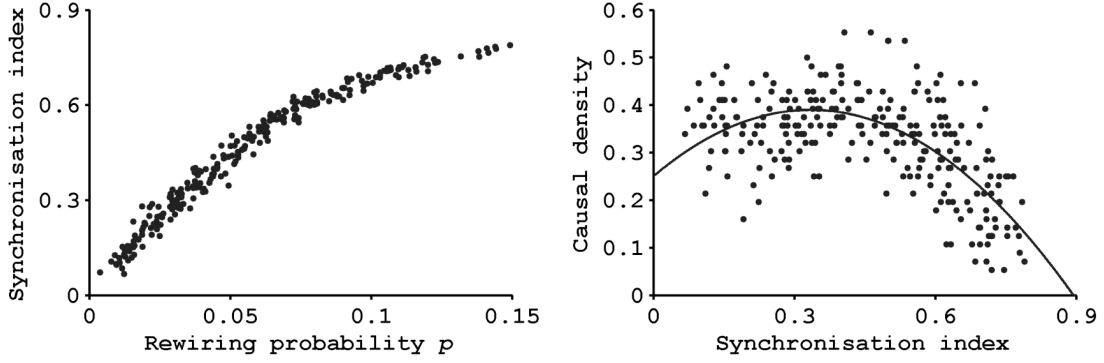

FIG. 6: The relationship between rewiring probability, synchronisation index, and causal density for the 8-cluster experiment. See text for details.

It is possible to quantify the phenomenon in question by assessing the level of global phase synchrony among the clusters. Let $X_1(t)$ and $X_2(t)$ be two scalar signals and let $\phi_1(t)$ and $\phi_2(t)$ respectively denote their instantaneous phases according to the Hilbert transform. Then, following Rosenblum, *et al.* [21], an estimate of the *phase synchrony* between $X_1(t)$ and $X_2(t)$ is given by

$$\Gamma = \langle \cos \Psi(t) \rangle^2 + \langle \sin \Psi(t) \rangle^2$$

where $\langle f(t) \rangle$ denotes the average of *f* over time, and

$$\Psi(t) = (\phi_1(t) - \phi_2(t)) \bmod 2\pi.$$

The *synchronisation index* among a system of variables $X_1...X_n$ may then be defined as the phase synchrony estimate $\Gamma$ averaged over all pairs of distinct variables $\langle X_i, X_j \rangle$.

As Fig. 6 (left) illustrates for the 8-cluster experiment, phase synchrony increases monotonically with rewiring probability, which accords with the above explanation. (Comparable results are reported by Masuda & Aihara [22] and Percha, *et al*. [23] for small-world networks of neurons with topologies different from that used here, and by Park, *et al*. [24] and Guan, *et al*. [20] for small-word networks of coupled oscillators with a topology similar to that used here.) Fig. 6 (right) also shows, while phase synchrony is prevalent in all complete runs, phase synchrony indices above 0.4 (corresponding roughly with rewiring probabilities > 0.05) are inversely correlated with causal density (cf: Fig. 3(a)), which again accords with the above explanation.

ACKNOWLEDGMENTS

The author gratefully acknowledges the use of the "brain connectivity" Matlab toolbox maintained by Olaf Sporns, as well as Anil Seth's "causal connectivity analysis" toolbox. Thanks also to Anil Seth for many fruitful discussions relating to this paper, and to one of the anonymous referees whose feedback was especially helpful

REFERENCES

1. D.S.Bassett and E.Bullmore, The Neuroscientist **12**, 512 (2006).
2. O.Sporns and J.D.Zwi, Neuroinformatics **2**, 145 (2004).

3. G.F.Striedter, *Principles of Brain Evolution* (Sinauer Associates, 2005).

4. Q.Wen and D.B.Chklovskii, PLoS Computational Biology **1**, 0617 (2006).

5. A.K.Seth, E.M.Izhikevich, G.N.Reeke and G.M.Edelman, Proc. Natl. Acad. Sci. USA **103**, 10799 (2006).

6. M.Shanahan, Consciousness and Cognition **17**, 288 (2008).

7. G.Tononi, G.M.Edelman and O.Sporns, Trends in Cognitive Sciences **2**, 474 (1998).

8. O.Sporns, G.Tononi and G.M.Edelman, Cerebral Cortex **10**, 127 (2000).

9. M.Rosenblum, A.Pikovsky, J.Kurths, C.Schäfer and P.A.Tass, in *Handbook of Biological Physics, Vol. 4: Neuro-Informatics and Neuromodelling*, edited by F.Moss and S.Gielen (Elsevier, 2001), Chapter 9, pp. 279–321.

10. A.Roxin, H.Riecke and S.A.Solla, Phys. Rev. Lett. **92**, 198101 (2004).

11. D.J.Watts and S.H.Strogatz, Nature **393**, 440 (1998).

12. P.Hagmann, L.Cammoun, X.Gigandet, R.Meuli, C.J.Honey, V.J.Wedeen and O.Sporns, PLoS Biology **6**, e159 (2008).

13. M.Girvan and M.E.J.Newman, Proc. Natl. Acad. Sci. USA **99**, 7821 (2002).

14. R.K.Pan and S.Sinha, arXiv:0802.3671v1 [physics.bio-ph], 2008 (unpublished).

15. E.M.Izhikevich, IEEE Trans. Neural Networks **14**, 1569 (2003).

16. M.D.Humphries, K.Gurney and T.J.Prescott, Proc. Roy. Soc. B **273**, 503 (2006).

17. A.K.Seth, Cognitive Neurodynamics **2**, 49 (2008).

18. C.W.J.Granger, Econometrica **37**, 424 (1969).

19. C.Buckley and S.Bullock, in *Advances in Artificial Life, Lecture Notes in Computer Science Vol. 4648*, edited by F.Almedia e Costa, L.M.Rocha, E.Costa, I.Harvey and A.Coutinho (Springer, 2007), pp. 986–995.

20. S.Guan, X.Wang, Y.-C.Lai and C.-H.Lai, Phys. Rev. E **77**, 046211 (2008).

21. M.Rosenblum, A.Pikovsky, J.Kurths, C.Schäfer and P.A.Tass, in *Handbook of Biological Physics, Vol. 4: Neuro-Informatics and Neuromodelling*, edited by F.Moss and S.Gielen (Elsevier, 2001), Chapter 9, pp. 279–321.

22. N.Masuda and K.Aihara, Biological Cybernetics **90**, 302 (2004).

23. B.Percha, R.Dzakpasu, M.Zochowski and J.Parent, Phys. Rev. E **72**, 031909 (2005).

24. K.Park, Y.-C. Lai , S.Gupte and J.-W.Kim, Chaos **16**, 015105 (2006).